\documentclass[%
 reprint,
 amsmath,amssymb,
prb,
]{revtex4-1}

\usepackage{graphicx}
\usepackage{dcolumn}
\usepackage{bm}

\usepackage[hypertex, linktocpage]{hyperref}[2003/11/30]
\usepackage{siunitx}
\usepackage{bm}
\usepackage{verbatim}
\usepackage[version=4]{mhchem}

\begin{document}

\preprint{APS/123-QED}

\title{Ferromagnetic resonance in the complex permeability of an \ce{Fe3O4} nanosuspension at radio and microwave frequencies}

\author{J. Dubreuil}
\author{J. S. Bobowski}
\email{jake.bobowski@ubc.ca}
\homepage{\\https://people.ok.ubc.ca/jbobowsk}
\affiliation{Department of Physics, University of British Columbia, Kelowna, British Columbia, V1V 1V7, Canada}


\date{\today}

\begin{abstract}
The complex permeability of an iron-oxide nanosuspension has been measured as a function of magnetic field strength at RF and microwave frequencies using a loop-gap resonator.  The particles were suspended in water and had an \SI[number-unit-product=\text{-}]{8}{\nano\meter} diameter \ce{Fe3O4} core that was coated by Dextran.  The real part of the permeability increased sharply beyond a frequency-dependent threshold value of the static magnetic field before saturating.  Just beyond this threshold field, there was a peak in the imaginary part of the permeability.  The permeability measurements, which exhibited features associated with ferromagnetic resonance, were used to determine the dependence of the microwave absorption on static magnetic field strength.  Using the absorption data, the $g$-factor of the nanosuspension was found to be \SI[separate-uncertainty = true, multi-part-units=single]{1.86\pm 0.07}{}.   
%
\end{abstract}

\maketitle

%


{\it Introduction}.\ Since the discovery of the metal-to-insulator transition in magnetite, \ce{Fe3O4} has been one of the most widely studied magnetic systems.\cite{Okamura:1932, Verwey:1939}  Upon cooling magnetite below the so-called Verwey transition temperautre $T_\mathrm{V}\approx\SI{125}{\kelvin}$, its resistivity abruptly increases by up to two orders of magnitude in the best single crystals.  As described in a review by Walz, there are numerous other anomalies in the physical properties of \ce{Fe3O4} at $T_\mathrm{V}$, such as steps in the magnetization, initial susceptibility, and thermal expansion, and a spike in the specific heat.  However, despite decades of intense research, there is not yet a consensus understanding of the microscopic mechanisms governing the Verwey transition\cite{Walz:2002} and it is still being actively investigated both experimentally\cite{Bartelt:2013, Ramos:2014, Taguchi:2015, Senn:2015, Randi:2016, Garcia:2016, Ma:2017, Borroni:2017} and theoretically\cite{Seo:2002, Jeng:2004, Craco:2006}.

Shortly after the discovery of the Verwey transition, Bickford observed resonant microwave absorption in \ce{Fe3O4} single crystals.\cite{Bickford:1950}  Based on  the theory of ferromagnetic resoance developed by Kittel,\cite{Kittel:1948, Kittel:1949} these data were used to determine the crystalline anisotropy and $g$-factor of magnetite.  More recently, there has been considerable interest in the magnetic properties of \ce{Fe3O4} thin films\cite{McGuigan:2008, Monti:2012, Nagata:2014} and nanoparticles\cite{Ni:2009, Jia:2010}.  \ce{Fe3O4} nanaoparticles, in particular, have attracted a lot of attention because of their potential use in biomedical,\cite{Dinali:2017} environmental,\cite{Xu:2012} and other applications.\cite{Hasany:2012}

In this paper, we describe our measurements of the complex permeability of a suspension of \SI[number-unit-product={\text{-}}]{8}{\nano\meter} diameter \ce{Fe3O4} particles as a function static magnetic field strength.  The measurements were made using a novel compact resonator whose resonant frequency could be tuned between \SI{500}{} and \SI{1300}{\mega\hertz}.  Many previous studies of the microwave properties of \ce{Fe3O4} nanoparticles have used less sensitive non-resonant techniques, often with larger diameter partilces (\SI{150}{} to \SI{300}{\nano\meter}).  Furthermore, in these studies the authors were primarily interested in the microwave absorption properties in zero static magnetic field.\cite{Ni:2009, Jia:2010}  Noginov {\it et al.}\ studied the electron magnetic resonance (EMR) signal in a suspension of \SI[number-unit-product={\text{-}}]{9}{\nano\meter} diameter \ce{Fe3O4} nanoparticles and identified features which marked quantum transitions.\cite{Noginov:2008}  Shankar {\it et al.}\ likewise studied magnetite nanoparticles that were \SI{9}{} to \SI{10}{\nano\meter} in size.  These authors investigated the temperature dependence of the field-cooled and zero-field cooled ferromagnetic resonance lineshapes at \SI{9.45}{\giga\hertz} and identified a spin glass transition at \SI{46}{\kelvin}.\cite{Shankar:2015}  The measurements that we present offer a detailed determination of the field dependence of the complex permeability of an \ce{Fe3O4} nanosuspension over a relatively wide range of RF frequencies. 

The organization of the paper is as follows: We first briefly introduce the the loop-gap resonator (LGR) and the measurement principle.  The experimental design is then described along with a discussion of its advantageous and limitations.  The experimental results are then presented and compared to complementary measurements on \ce{Fe3O4} bulk single crystals and thin films.  Finally, the key findings are briefly summarized.

{\it Lop-gap resonator}.\ The measurements presetned in this paper were made using a loop-gap resonator (LGR).  In its most basic form, the LGR is made by cutting a narrow slit along the length of a hollow conducting tube.\cite{Hardy:1981,Froncisz:1982}   The effective capacitance of the gap and inductance of the bore determine the resontant frequency $f_0$ (typically between \SI{100}{\mega\hertz} and \SI{2}{\giga\hertz}).  The quality factor or $Q$ of the resonance is determined by electromagnetic skin depth and radiative losses.\cite{Bobowski:2013}  The radiative losses associated with conventional cylindrical LGRs can be suppressed using a toroidal geometry which completely confines the magnetic flux within the bore of the resonator.\cite{Bobowski:2016}  The advantages of the LGR include resonator dimensions that are many times smaller than the free-space wavelength of the resonant frequency and good isolation between RF electric and magnetic fields. 

\begin{figure*}
\begin{tabular}{rl}
(a)~\includegraphics[width=7 cm, keepaspectratio]{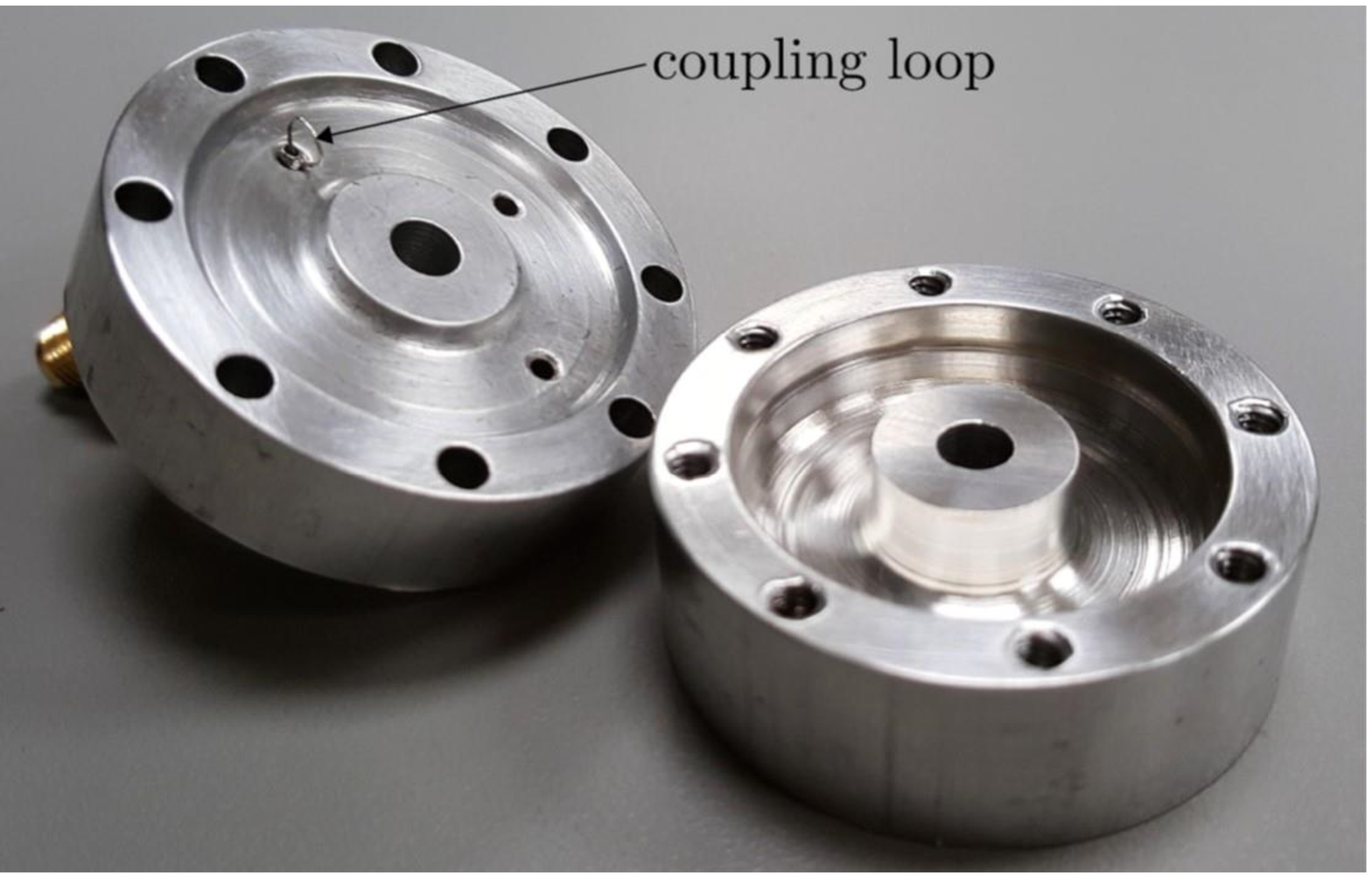}\quad~ &
~\quad(b)\includegraphics[width=8.5 cm, keepaspectratio]{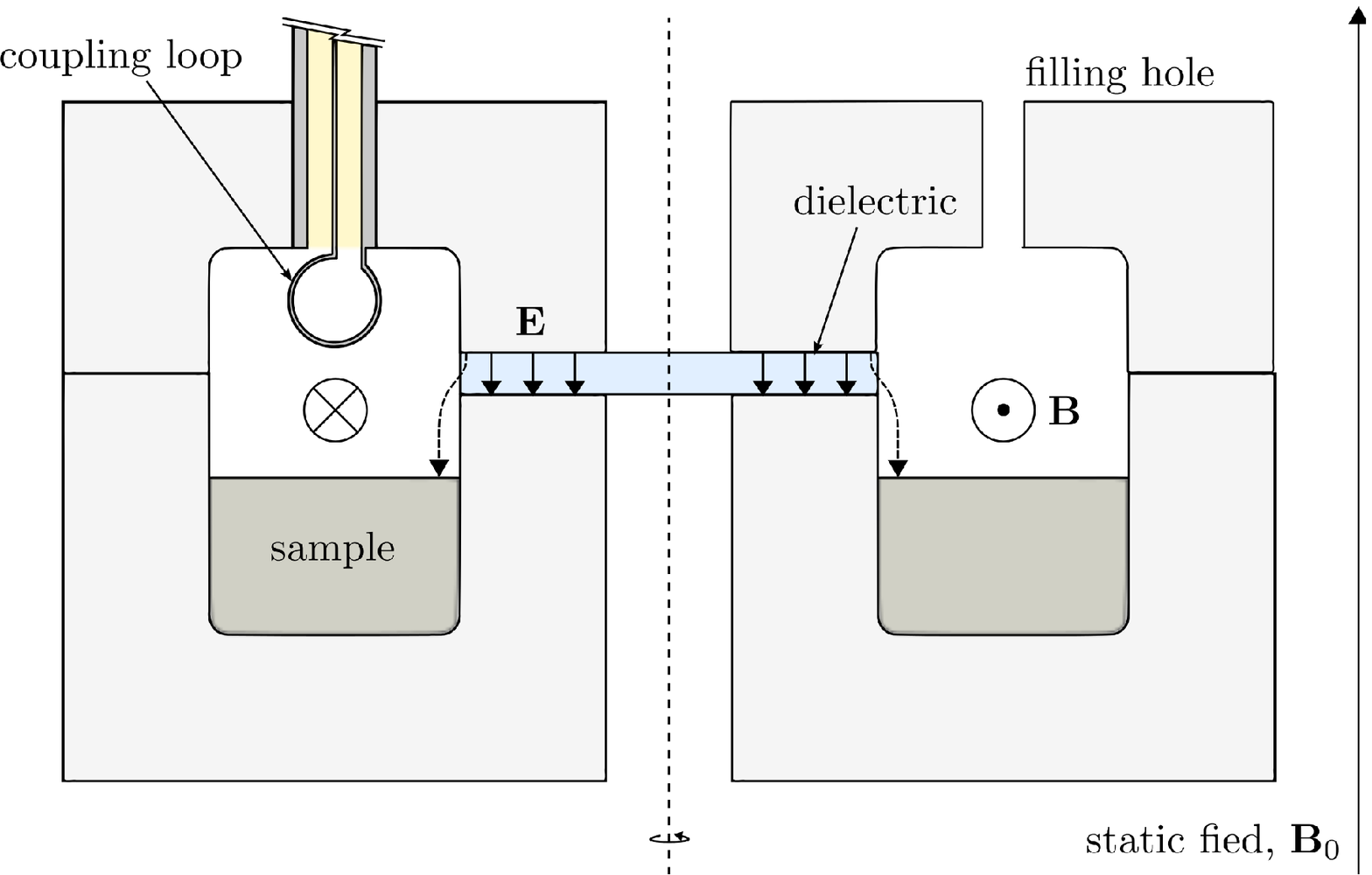}
\end{tabular}
\caption{\label{fig:LGR}(a) Photograph of the two halves of the toroidal LGR with an asymmetrically-placed gap.  A coupling loop passes through a small hole in the top half of the resonator.  The outside diameter of the LGR is \SI{3.8}{\centi\meter} (\SI{1.5}{inches}).  (b) Schematic diagram of the cross-section of the toroidal LGR.  The resonant frequency is set by a low-loss dielectric in the gap of the resonator.  The magnetic nanosupension partially fills the bore of the resonator.  A uniform static magnetic field $\mathbf{B}_0$ is applied perpendicularly to the RF magnetic field $\mathbf{B}$.}
\end{figure*}

Inserting an insulator into the gap of a LGR allows one to determine the material's complex permittivity.  The shift in the resonant frequency is determined by the real part of the permittivity while the imaginary part lowers the quality factor of the resonance.\cite{Bobowski:2013, Bobowski:2017}  In an analogous way, the complex permeability of a magnetic material loaded in the bore of a LGR can likewise be determined from the changes in $f_0$ and $Q$.\cite{Bobowski:2015, Bobowski:2018}  This paper describes the use of a toroidal LGR to measure the change in the complex permeability of an \ce{Fe3O4} nanosuspension as the strength of an applied static magnetic field is varied.   

\begin{figure*}
\includegraphics[width=14 cm, keepaspectratio]{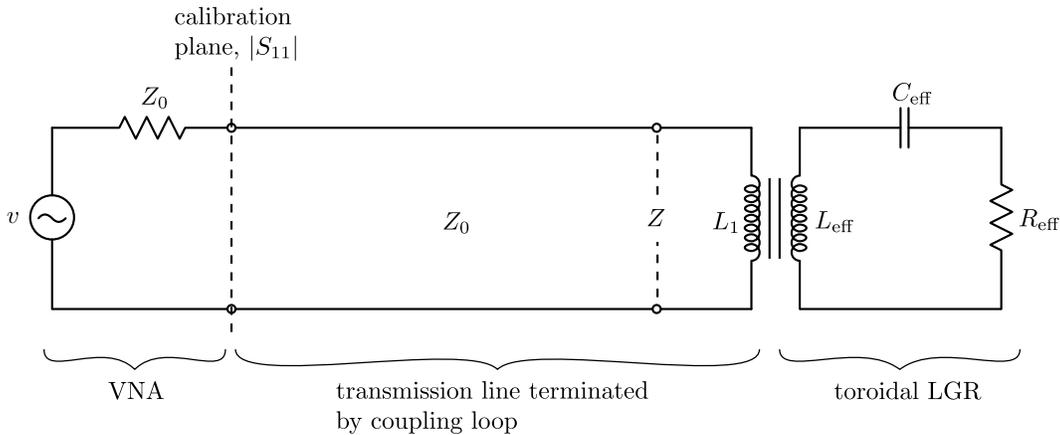}
\caption{\label{fig:circuit}Complete equivalent circuit of the toroidal LGR coupled to a length of transmission line terminated by a coupling loop.  A VNA is used to measure the magnitude of the reflection coefficient $\left\vert S_{11}\right\vert$ at a calibration plane established at the opposite end of the transmission line.}
\end{figure*}

{\it Experimental method}.\ Because a new experimental method was designed specifically for this measurement, we briefly outline some of the details.  Figure~\ref{fig:LGR} shows a photograph of the aluminum toroidal LGR used in our experiments along with a schematic diagram of its cross-section.  An ac magnetic flux is introduced into the bore of the LGR using a coupling loop at the end of a short section of coaxial transmission line.  The opposite end of the transmission line is connected to one port of a vector network analyzer (VNA).  The VNA supplies a signal and measures what fraction of the incident signal is reflected back as a function of frequency.  
A full equivalent-circuit model of the experimental setup in shown in Fig.~\ref{fig:circuit}.\cite{Rinard:1993}

As shown in Fig.~\ref{fig:LGR}(b), the \ce{Fe3O4} nanosuspension only partially fills the bore of the LGR.  By the symmetry of the toroidal geometry, the current density on the inner bore wall is uniform.  That is, the current density is the same in the filled and unfilled sections of the bore.  On the other hand, the magnetic flux within cross-sections of the filled and unfilled regions of the bore are certainly different.  These observations suggest that the effective inductance of the LGR bore can be modeled as the series combination \mbox{$L_\mathrm{eff}=L_1+\mu_\mathrm{r} L_2$} where \mbox{$L_1=\left(1-\eta\right)L_0$} is the inductance of the unfilled region, \mbox{$L_2=\mu_\mathrm{r}\eta L_0$} is the inductance of the filled region, $\eta$ is the filling fraction, $L_0$ is the inductance of the bore when $\eta=0$, and \mbox{$\mu_\mathrm{r}=\mu^\prime-j\mu^{\prime\prime}$} is the complex relative permeability of the nanosuspension (\mbox{$j=\sqrt{-1}$}).\cite{Bobowski:2018}  Therefore, the effective inductance shown Fig.~\ref{fig:circuit} is complex and can be expressed as
\begin{equation}
L_\mathrm{eff}=\left\{\left[1+\eta\left(\mu^\prime-1\right)\right]-j\eta\mu^{\prime\prime}\right\}L_0.
\end{equation}
As a result, the impedance \mbox{$j\omega L_\mathrm{eff}$} has a resitive term given by \mbox{$\eta\mu^{\prime\prime}\omega L_0$} that combines in series with $R_\mathrm{eff}$.  Thus, a non-zero value of $\mu^{\prime\prime}$ enhances the net resistance and suppresses the resonator $Q$.

Figure~\ref{fig:LGR}(b) shows that the gap of the LGR has been placed asymmetrically towards to the top of the resonator.  This was done to provide more sample space for the nanosuspension and also to limit, as much as possible, fringing electric fields emanating from the gap from reaching the sample. The figure also shows a low-loss dielectric filling the gap of the resonator.  The dielectric is used to adjust the resonance frequency of the LGR.  Changing the dielectric allowed us to tune the resonant frequency to values between \SI{500}{} and \SI{1300}{\mega\hertz} with\-out significantly reducing $Q$.  The dielectrics used were \ce{Al2O3}, \ce{TiO2} ([100] and [001] substrates), and \ce{SrTiO2}.\footnote{All dielectric substrates were purchased from MTI Corporation (Richmond, CA).}  The $C_\mathrm{eff}$ parameter in the circuit model is predominantly set by the dimensions of the slit in the LGR and the dielectric filling the gap.  However, fringing electric fields that reach the sample also make a small contribution to $C_\mathrm{eff}$ such that \mbox{$C_\mathrm{eff}=C_0+\varepsilon_\mathrm{r}C_\mathrm{f}$} where \mbox{$C_0+C_\mathrm{f}$} is the capacitance when the gap is filled with a low-loss dielectric and the resonator bore is empty and \mbox{$\varepsilon_\mathrm{r}=\varepsilon^\prime-j\varepsilon^{\prime\prime}$} is the complex permittivity of the material that partially fills the resonator bore.\cite{Stuchly:1980, Bobowski:2012}

The remaining circuit-model parameters are \mbox{$R_\mathrm{eff}\approx R_0\sqrt{f/f_0}$} where $R_0$ is the effective resistance at $f_0$,\cite{Bobowski:2016} $L_1$ which is the inductance of the coupling loop, and \mbox{$Z_0=\SI{50}{\ohm}$} which is both the output impedance of the VNA and the characteristic impedance of the transmission line.

The circuit in Fig.~\ref{fig:circuit} can be analyzed to determine the impedance $Z$ of the inductively-coupled LGR.\cite{Rinard:1993, Bobowski:2018}  The magnitude of reflection coefficient at the calibration plane of the VNA is then given by
\begin{equation}
\left\vert S_{11}\right\vert=\left\vert\frac{Z-Z_0}{Z+Z_0}\right\vert.
\end{equation}
A total of four successive measurements are required to determine all of the circuit-model parameters:\cite{Bobowski:2018} (1) A measurement of $S_{11}$  when the LGR is not in place determines $L_1$.  (2) A measurement of $\left\vert S_{11}\right\vert$ when the bore of the resonator is empty determines $f_0$, $Q$, and $M_0^2/R_0$ where $M_0$ is the mutual inductance between $L_1$ and $L_0$. (3) A measurement of $\left\vert S_{11}\right\vert$ when the bore is filled with deionized water, assuming a known filling fraction $\eta$ and $\varepsilon_\mathrm{r}$ of water, is used to determine the ratio $C_\mathrm{f}/C_0$. (4) Finally, a measurement of $\left\vert S_{11}\right\vert$ when the bore is partially filled with the \ce{Fe3O4} nanosuspension is used to determine $\mu^\prime$ and $\mu^{\prime\prime}$.  

The final measurement assumes that the complex permittivity of the nanosuspension is known.  Because we have not precisely determined $\varepsilon_\mathrm{r}$ of the nanosuspension, our measurements cannot be used to extract the absolute complex permeability of our sample.  However, the change in $\left\vert S_{11}\right\vert$ in the presence of an applied static magnetic field is very insensitive to the values of both $\varepsilon_\mathrm{r}$ and $C_\mathrm{f}/C_0$.  Therefore, our measurements can precisely determine the dependence of the nanosuspension's complex permeability \mbox{$\left(\mu^\prime-j\mu^{\prime\prime}\right)$} on the strength of the static field.  In our experiments, the static magnetic field was generated using a solenoid wound using \SI[number-unit-product=\text{-}]{18}{AWG} copper wire.  The solenoid had a bore diameter of \SI{7.6}{\centi\meter} and produced a magnetic field of \SI{50}{\milli\tesla} when supplied with \SI{4}{\ampere} of current.

{\it Nanosuspension}.\ The nanosuspension used in our experiments, called Molday ION, is commercially available from BioPhysics Assay Laboratory, Inc.\cite{BioPAL}  It is a suspension of Dextran-coated \ce{Fe3O4} nanoparticles in distilled water.  The mean diameter of the \ce{Fe3O4} core is \SI{8}{\nano\meter} and Molday ION contains \SI{10}{\milli\gram} of \ce{Fe} per milli-liter of suspension.\cite{Molday}  The Dextran coating is used to suppress amalagamation of the \ce{Fe3O4} nanoparticles and to increase the biocompatibility of the suspension for use in biomedical applications.  Molday ION is primarily used as an MRI contrast agent (see, for example, Refs.~\onlinecite{Hyodo:2009, Perles:2012, Hitchens:2015}).  All of our permeability measurements were made using \SI{2}{\milli\liter} of the Molday ION suspension.

\begin{figure}[t]
\begin{tabular}{lr}
\includegraphics[height=8.5cm, keepaspectratio]{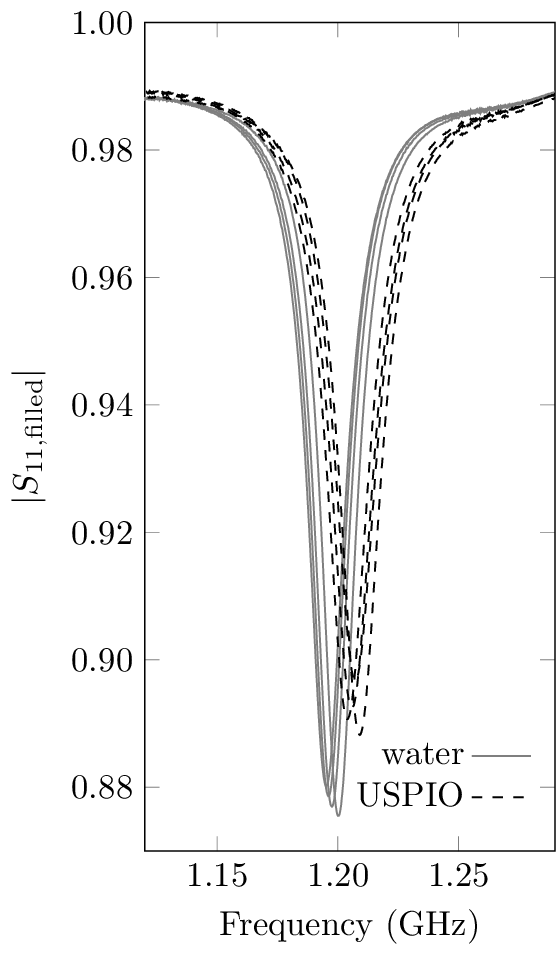} & \includegraphics[height=8.5cm, keepaspectratio]{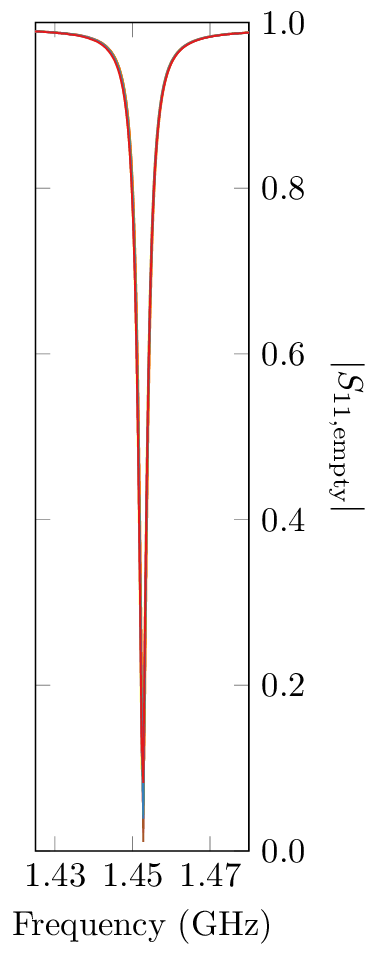}
\end{tabular}
\caption{\label{fig:resonances}The right-hand plot shows the resonance of the toroidal LGR when its bore was empty.  There are eight nearly-identical measurements of $\left\vert S_{11,\mathrm{empty}}\right\vert$ shown in the plot.  The left-had plot shows measured reflection coefficient when the bore of the resonator was filled with \SI{2}{\milli\liter} of water (solid lines) and \SI{2}{\milli\liter} of the Fe$_3$O$_4$ nanosuspension (dashed lines).  There are four measurements each for a water-filled and nanosuspension-filled LGR.  All of these data were obtained with an \ce{Al2O3} plate in the gap of the resonator.}
\end{figure}

{\it Experimental results}.\ Figure~\ref{fig:resonances} shows the measured $\left\vert S_{11}\right\vert$ when the resonator bore was empty, filled with \SI{2}{\milli\liter} of water, and filled with \SI{2}{\milli\liter} of the \ce{Fe3O4} nanosuspension.  For these measurements, the LGR gap was filled with a single-crystal \ce{Al2O3} plate.  The LGR was also equipped with a platinum resistance thermometer.  All $\left\vert S_{11}\right\vert$ measurements reported in this paper were taken at a LGR temperature of \SI[separate-uncertainty = true, multi-part-units=single]{25.0\pm 0.1}{\celsius}. 

There are eight nearly-indistinguishable plots of $\left\vert S_{11}\right\vert$ measured while the LGR bore was empty.  Between each measurement, the two halves of the LGR (see Fig~\ref{fig:LGR}(a)) were completely separated.  By adjusting the torque applied to the bolts used to assemble the LGR while watching the VNA display, it was possible to very reproducibly set the empty-bore resonant frequency.

The solid lines in the left-hand plot show $\left\vert S_{11}\right\vert$ after adding \SI{2}{\milli\liter} of deionized water to the resonator bore.   Adding the water to the bore reduced the resonant frequency by about 17\%.  This effect is due to fringing electric fields from the gap of the LGR that extend into the water. Water has a relatively large dielectric constant ($\varepsilon_\mathrm{r}\approx 78$ at \SI{25}{\celsius} and low frequencies) such that the contribution from $\varepsilon_\mathrm{r}C_\mathrm{f}$ becomes non-negligible.  Four measurements of $\left\vert S_{11}\right\vert$ for the water-filled resonator are shown.  Between each measurement, the water was completely drained from the bore of the resonator and the resonator was completely disassembled.  The data show that the $\left\vert S_{11}\right\vert$ measurements are reproducible.

The dashed lines in Fig.~\ref{fig:resonances} show $\left\vert S_{11}\right\vert$ measured when \SI{2}{\milli\liter} of the \ce{Fe3O4} is added to the LGR bore.  Once again, repeated measurements gave consistent results.  Relative to the water measurements, the nanosuspension data show a slight increase in the resonant frequency.  This shift in frequency is not due to a magnetic effect.  Rather, it is caused by a nanosuspension dielectric constant that is slightly less than that of water.  The volume of water displaced by the nanoparticles results in a lower effective dielectric constant of the suspension and shifts the resonance to a higher frequency.  This effect completes with, and dominates, the expected paramagnetic permeability of the nanosuspension which would tend to lower the resonant frequency relative to water.  For this reason, the absolute permeability of the nanosuspension cannot be determined without precise knowledge of the permittivity of both water and the nanosuspension at the measurement frequency.  However, in an applied magnetic field, only changes in nanosuspension's permeability can result in a change to $\left\vert S_{11}\right\vert$.  Therefore, the observed magnetic field dependence of $\left\vert S_{11}\right\vert$ can be used to determine how $\mu^\prime$ and $\mu^{\prime\prime}$ deviate from their zero-field values.

\begin{figure}[t]
\includegraphics[width=0.9\columnwidth, keepaspectratio]{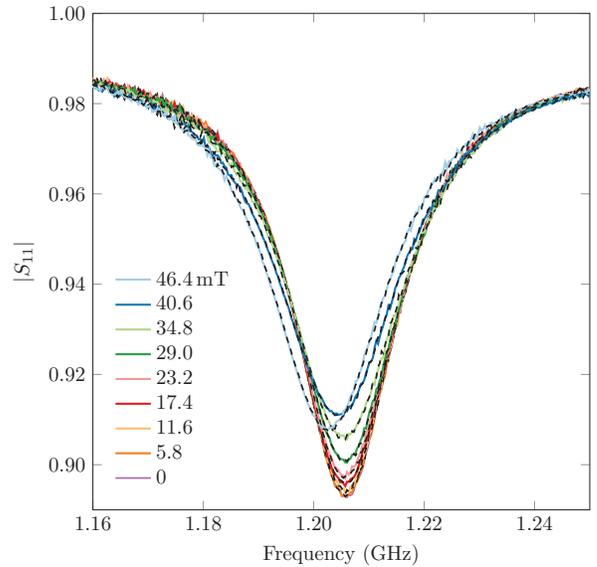}
\caption{\label{fig:Bsweep}Measurements of $\left\vert S_{11}\right\vert$ with the bore of the LGR partially filled with the nanosuspension at different static magnetic field strengths.  The solid curves were obtained while reducing the field strength from \SI{50}{\milli\tesla} to zero and the dashed curves were taken while increasing the strength of the static field.  For these measurements the gap of the LGR was filled with an \ce{Al2O3} dielectric.}
\end{figure}

Figure~\ref{fig:Bsweep} shows the evolution $\left\vert S_{11}\right\vert$ as the static magnetic field strength applied to the nanosuspension-filled LGR is varied.
The data show the full range of magnetic field strengths explored, but only a subset of the values tested.  The solid lines correspond $\left\vert S_{11}\right\vert$ measured while sweeping from high field to low field and the dashed lines correspond to a low-to-high field sweep.  The overlap of the solid and dashed lines confirms that there was no appreciable temperature drift of the LGR/sample during the magnetic field sweep.  We did identical magnetic field sweeps when the bore of the resonator was empty and filled with \SI{2}{\milli\liter} of water.  In both cases, $\left\vert S_{11}\right\vert$ did not deviate from its zero-field response at any of the magnetic field strengths used in our experiments.

Before discussing the quantitative results, we first qualitatively interpret the $\left\vert S_{11}\right\vert$ data of Fig.~\ref{fig:Bsweep}.  At the lowest static field strengths, the resonant frequency is approximately constant which implies a constant $\mu^\prime$.  After reaching a threshold field of about \SI{30}{\milli\tesla}, the $\left\vert S_{11}\right\vert$ resonant frequency starts to decrease which corresponds to an increasing $\mu^\prime$.  The depth of the $\left\vert S_{11}\right\vert$ dip can be used to track the changes in $Q$ -- the greater the depth, the higher the quality factor.  Figure~\ref{fig:Bsweep}, shows that $Q$ initially decreases with increasing field before reaching a minimum and then increasing.  These data suggest a $\mu^{\prime\prime}$ that increases, peaks, and then decreases as a function of the applied static field strength.  The data imply that $\mu^{\prime\prime}$ peaks at static field strength near \SI{40}{\milli\tesla}.

\begin{figure*}
\begin{tabular}{lr}
(a)\includegraphics[height=7.25cm, keepaspectratio]{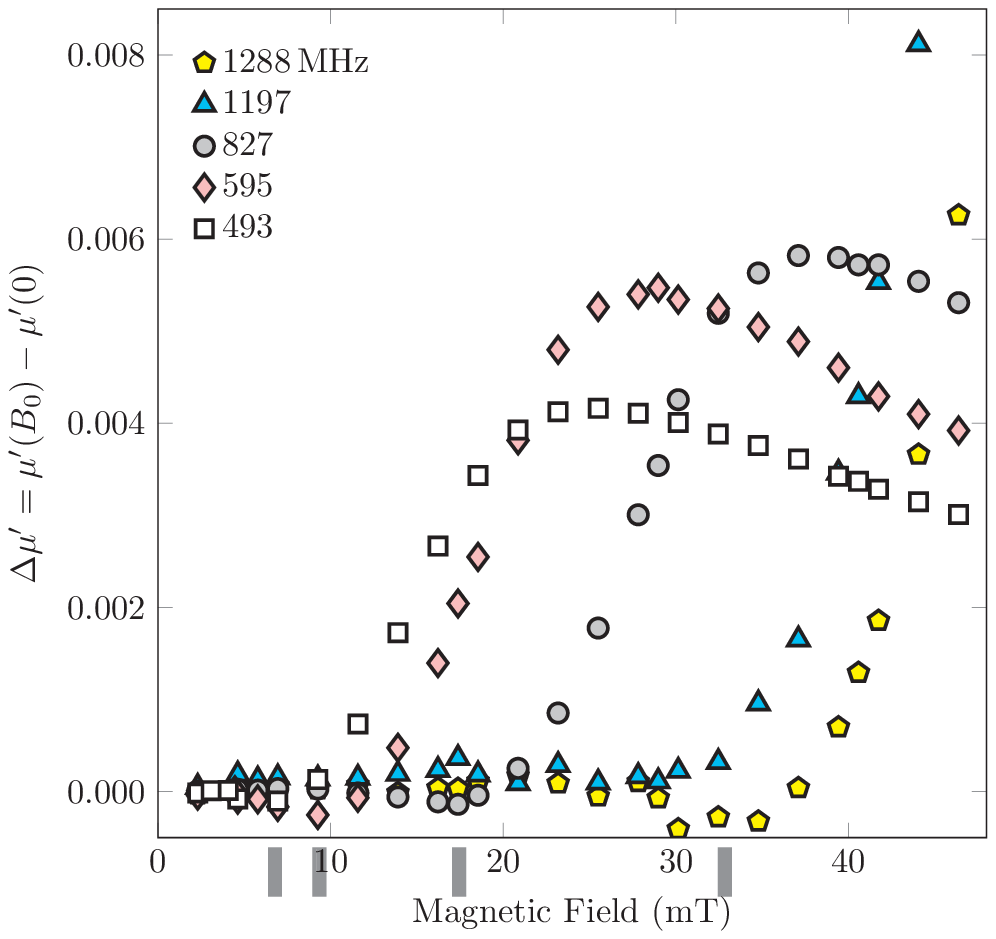}\quad~ & ~\quad(b)\includegraphics[height=7.25cm, keepaspectratio]{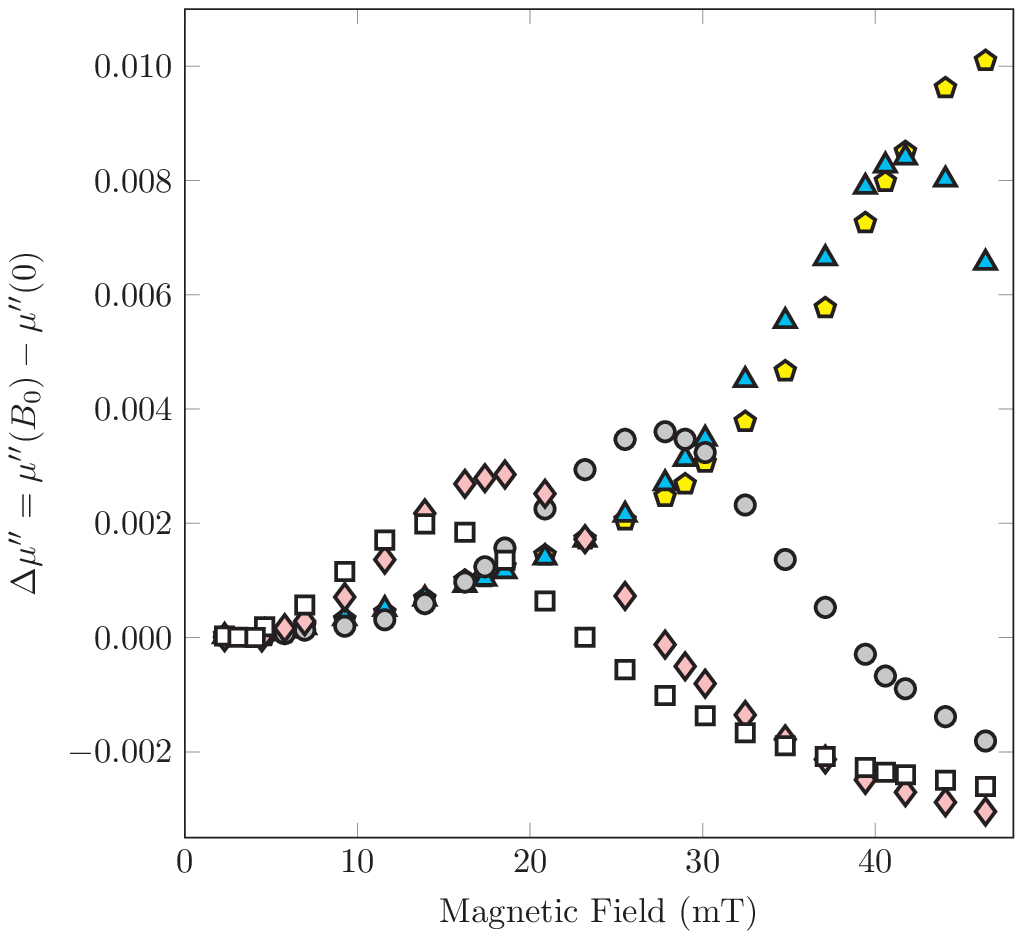}
\end{tabular}
\caption{\label{fig:permeability}The real and imaginary parts of the change in permeability of the Fe$_3$O$_4$ nanosuspension verses static magnetic field strength. (a) The real part of the permeability, initially constant, rises sharply beyond a threshold value of the static field strength and then saturates.  The gray markers highlight minima in $\Delta\mu^\prime$ that occur just before the sharp increase. (b)  The imaginary part of the permeability peaks at a frequency-dependent value of the magnetic field strength.}
\end{figure*}

The $\left\vert S_{11}\right\vert$ data of Fig.~\ref{fig:Bsweep} were fit to the equivalent circuit model and the extracted changes in $\mu^\prime$ and $\mu^{\prime\prime}$ versus $B_0$ are shown using triangular data points in Fig.~\ref{fig:permeability}.
As anticipated, \mbox{$\Delta\mu^\prime\equiv \mu^\prime\left(B_0\right)-\mu^\prime\left(0\right)$} is initially flat and then increases sharply after exceeding a threshold value of $B_0$.  After the sharp increase, the $\Delta\mu^\prime$ data plateau. \mbox{$\Delta\mu^{\prime\prime}\equiv \mu^{\prime\prime}\left(B_0\right)-\mu^{\prime\prime}\left(0\right)$} goes through a broad peak as a function of magnetic field with the peak of $\Delta\mu^{\prime\prime}$ located at the predicted value of $B_0$.

Our measurements of $\Delta\mu^\prime$ and $\Delta\mu^{\prime\prime}$ were done at a total of five different frequencies.  The LGR resonant frequency was tuned by changing the low-loss dielectric in the gap.  Figure~\ref{fig:permeability} shows all five data sets for both $\Delta\mu^\prime$ and $\Delta\mu^{\prime\prime}$.  The real part of the permeability remains constant up until a frequency-dependent threshold magnetic field after which $\Delta\mu^\prime$ rapidly increases.  For the three lowest measurement frequencies, $\Delta\mu^\prime$ saturates after the sharp rise.  The saturation region in the two highest-frequency measurements was not observed because we could not apply sufficient magnetic fields using our solenoid.  The $\Delta\mu^{\prime\prime}$ datasets all initially increase before reaching a peak value and then falling off as $B_0$ increases.  At the highest magnetic fields, we find \mbox{$\Delta\mu^{\prime\prime}<0$} in the three lowest-frequency datasets .  Extrapolating the $\Delta\mu^{\prime\prime}$ tails out to very high field strengths, where one expects $\mu^{\prime\prime}\to 0$, would be one way of estimating the zero-field values of $\mu^{\prime\prime}$. 

The $\Delta\mu_\mathrm{r}$ data have been measured with reasonably high resolution.  Given that the \ce{Fe} content in the nanosuspension is relatively dilute, the absolute permeability of the sample is expected to be close to one.  Therefore, we are measuring changes in the permeability with a precision that is greater than \SI{0.01}{\percent}.  The error bars in the $\Delta\mu^\prime$ measurements are approximately equal to the size of the data points and they are even smaller in the $\Delta\mu^{\prime\prime}$ measurements.   This precision has allowed us to identify a small but distinct feature in the $\Delta\mu^\prime$ data.  Just before the sharp increase, $\Delta\mu^\prime$ passes through a minimum.  These minima are clearly visible in all but the \SI{1197}{\mega\hertz} data and have be identified using rectangular markers below the magnetic field axis in Fig.~\ref{fig:permeability}(a).  Although we have not identified the source of the minima in $\Delta\mu^\prime$, they could be related to the minimum Bickford observed in his original microwave absorption experiments on bulk \ce{Fe3O4} crystals.\cite{Bickford:1950, Yager:1949}  

Energy dissipation, or absorption, in a ferromagnetic material is proportional to\cite{Yager:1949}
\begin{equation}
E_\mathrm{d}\propto\left[\sqrt{\left(\mu^\prime\right)^2+\left(\mu^{\prime\prime}\right)^2}+\mu^{\prime\prime}\right]^{-1/2}.
\end{equation}
For the relatively dilute \ce{Fe3O4} nanosuspension, the real and imaginary parts of the zero-field permeability can be written as \mbox{$\mu^\prime(0)=1+\delta_1$} and \mbox{$\mu^{\prime\prime}(0)=\delta_2$}, where \mbox{$\delta_1, \delta_2\ll 1$}.  Therefore, the components of permeability in a static magnetic field are given by \mbox{$\mu^\prime=1+\delta_1+\Delta\mu^\prime$} and \mbox{$\mu^{\prime\prime}=\delta_2+\Delta\mu^{\prime\prime}$} such that, to within a very good approximation
\begin{equation}
E_\mathrm{d}\propto 1+\frac{\delta_1+\delta_2}{2}+\frac{\Delta\mu^\prime+\Delta\mu^{\prime\prime}}{2}.
\end{equation}
The static field dependence of the energy dissipation is, therefore, simply determined from the sum \mbox{$\Delta\mu^\prime+\Delta\mu^{\prime\prime}$}.

\begin{figure*}
\begin{tabular}{lr}
(a)\includegraphics[width=0.9\columnwidth, keepaspectratio]{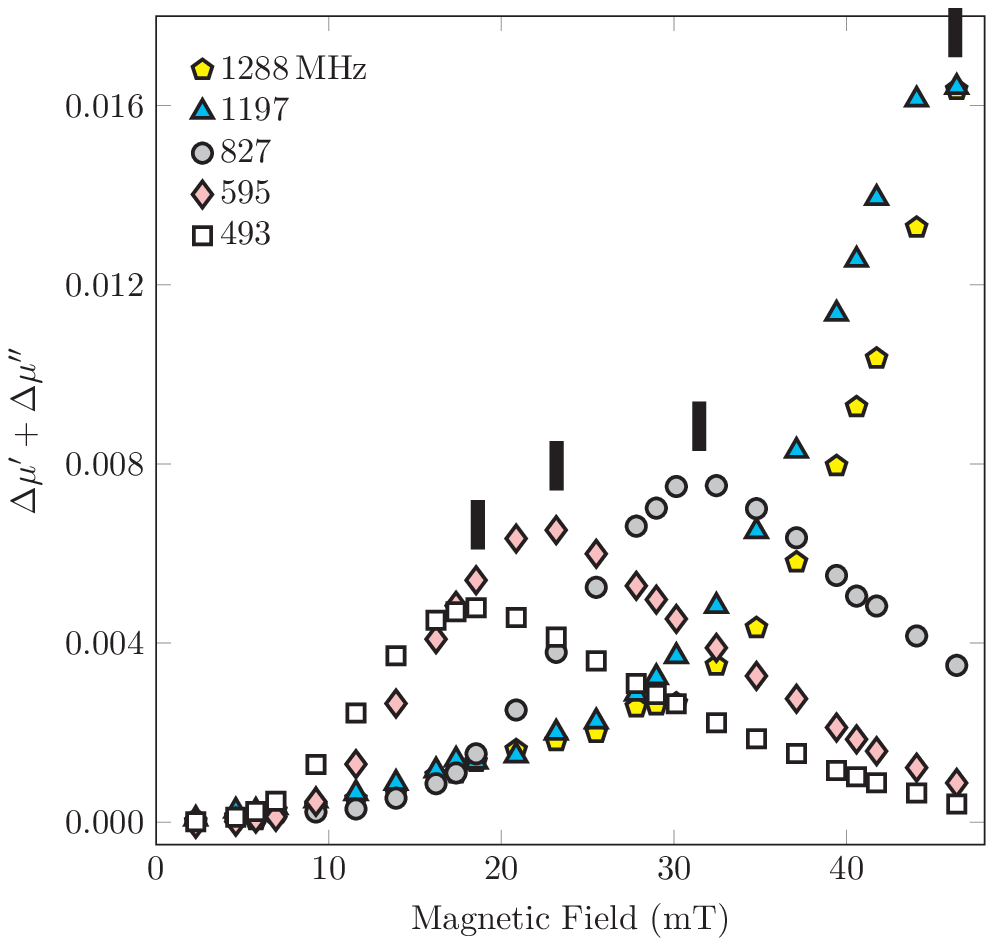}\quad~ & ~\quad(b)\includegraphics[width=0.9\columnwidth, keepaspectratio]{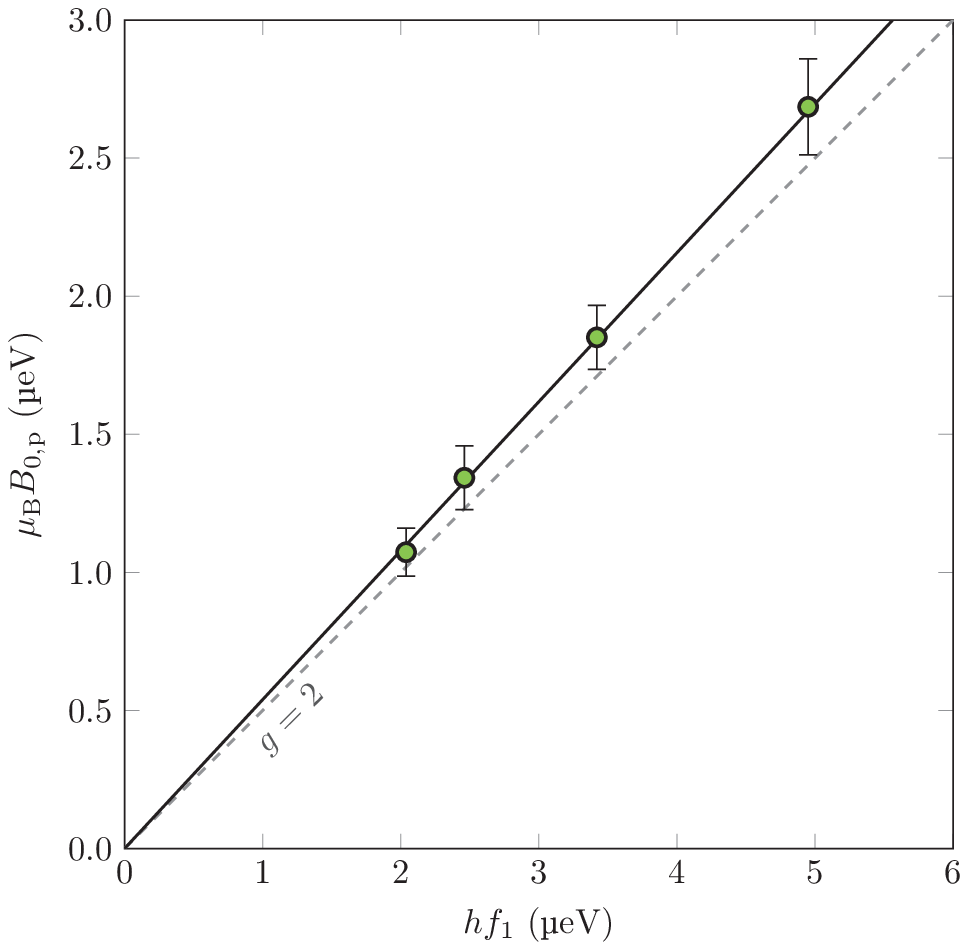}
\end{tabular}
\caption{\label{fig:energy}(a) Plot of the magnetic field dependence of the microwave absorption by the \ce{Fe3O4} nanosuspension at the five measurement frequencies.  (b) Plot of $\mu_\mathrm{B}B_\mathrm{0,p}$ as a function of $hf_1$.  $B_\mathrm{0,p}$ is the static field strength at which \mbox{$\Delta\mu^\prime+\Delta\mu^{\prime\prime}$} peaks and $f_1$ is the resonance frequency of the LGR when partially filled with the nanosuspension.  The data were fit to a straight line with zero intercept.  The dashed line has a slope of \SI{0.5}{} and corresponds to $g=2$.}
\end{figure*}

A plot of \mbox{$\Delta\mu^\prime+\Delta\mu^{\prime\prime}$} as a function of static magnetic field strength is shown in Fig.~\ref{fig:energy}(a).  
A broad peak in the microwave absorption is clearly visible in all but the \SI{1288}{\mega\hertz} dataset. Rectangular markers in the figure identify the values of the static field $B_\mathrm{0,p}$ at which the absorption peaks.  The plot in Fig.~\ref{fig:energy}(b) shows that $B_\mathrm{0,p}$ varies linearly with measurement frequency $f_1$.  Here, $f_1$ represents the resonant frequency of the LGR when its bore is filled with \SI{2}{\milli\liter} of the nanosuspension.  Assuming $g\mu_\mathrm{B}B_\mathrm{0,p}=hf_1$, where $\mu_\mathrm{B}$ is the Bohr magneton and $h$ is Planck's constant, the data have been fit to a straight line with zero intercept so as to extract a best-fit value for the $g$-factor.  The slope of the best-fit line is \SI[separate-uncertainty = true, multi-part-units=single]{0.539\pm 0.019}{} which corresponds to \mbox{$g=\SI[separate-uncertainty = true, multi-part-units=single]{1.86\pm 0.07}{}$}.  Microwave experiments on thin films of \ce{Fe3O4} have also reported values of $g$ that are less than two,\cite{Nagata:2014} whereas experiments on bulk single-crystals of magnetite find $g=2.12$ at room temperature.\cite{Bickford:1950} 

{\it Summary}.\ There has been growing interest in probing the magnetic properties of thin films, nanoparticles, and exotic nanostructures.\cite{Awschalom:1992, Wirth:1998, McGuigan:2008, Ni:2009, Jia:2010, Monti:2012, Liu:2013, Nagata:2014, Zhang:2014, Liu:2016, Zhang:2018} Using a LGR, we have developed a sensitive experimental technique to make precision measurements of the permeability of a dilute \ce{Fe3O4} nanosuspension as the strength of a static magnetic field was varied.  The LGR dimensions can be many times smaller than the free-space wavelength of the resonant frequency which allows one to work at low frequencies with relatively small sample volumes.  By placing low-loss dielectrics in the gap of the resonator, the measurement frequency was tuned from \SI{500}{} to \SI{1300}{\mega\hertz}.  

The real part of the nanosuspension's permeability was observed to increase sharply after exceeding a frequency-dependent threshold magnetic field.  $\Delta\mu^\prime$ then reached saturation at higher magnetic fields.  At four out of the five measurement frequencies, a small dip in $\Delta\mu^\prime$ was observed just before the onset of the sharp rise.  The imaginary part of the permeability exhibited a broad peak as a function of the applied static magnetic field strength. 

The $\Delta\mu^\prime$ and $\Delta\mu^{\prime\prime}$ measurements were used to extract the magnetic field dependence of the nanosuspension's microwave absorption.  The magnetic field $B_\mathrm{0,p}$ at the absorption peak was plotted as a function of the measurement frequency.  The resulting linear relationship, characteristic of a ferromagnetic resonance, yielded a $g$-factor of \SI[separate-uncertainty = true, multi-part-units=single]{1.86\pm 0.07}{}.

\begin{acknowledgments}
We wish to acknowledge the support of Thomas Johnson who generously provided access to an Agilent E5061A VNA. \end{acknowledgments}

\nocite{*}
\bibliography{J_Magn_Magn_Mater_Bobowski}

\end{document}